\documentclass[twocolumn,showpacs,showkeys]{revtex4}
\usepackage{graphicx}
\input{psfig.sty}

\newcommand{\bastar}{\begin{eqnarray*}}
\newcommand{\eastar}{\end{eqnarray*}}
\newskip\humongous \humongous=0pt plus 1000pt minus 1000pt

\newif\ifdtup

\relax
\newcommand{\be}{\begin{equation}}
\newcommand{\ee}{\end{equation}}
\newcommand{\bea}{\begin{eqnarray}}
\newcommand{\eea}{\end{eqnarray}}

\newcommand{\pro}{\partial}
\newcommand{\n}{\hat n}

\newcommand{\B}{{\vec B}}

\newcommand{\dfrac}{\displaystyle\frac}
\newcommand{\ba}{\begin{array}}
\newcommand{\ea}{\end{array}}

\newcommand{\nn}{\nonumber}
\newcommand{\hn}{\hat n}
\begin{document}
\title{Knots in Condensed Matters}
\author{Y. M. Cho}
\email{ymcho@yongmin.snu.ac.kr}
\affiliation{Center for Theoretical Physics and School of Physics \\
College of Natural Sciences, 
Seoul National University,
Seoul 151-742, Korea  \\
}
\begin{abstract}
~~~~~We propose two types of topologically stable knot solitons 
in condensed matters, one in two-component
Bose-Einstein condensates and one in two-gap superconductors. 
We identify the knot in Bose-Einstein condensates as
a twisted vorticity flux ring and the knot in two-gap superconductors
as a twisted magnetic flux ring.
In both cases we show that there is a remarkable interplay between
topology and dynamics which transforms the topologcal stability
to the dynamical stability, and vise versa.  
We discuss how these knots can be constructed in the spin-1/2 
condensate of $^{87}{\rm Rb}$ atoms and in two-gap superconductor
of $MgB_2$.
\end{abstract}
\maketitle

\section{Introduction}

Topological objects (monopoles, vortices, skyrmions, etc.) 
have played an increasingly important role
in physics \cite{dirac,abri}. In particular finite energy topological
objects have been widely studied in theoretical physics
\cite{skyrme,thooft}. A recent addition to this family of finite
energy solitons has been the knots \cite{fadd1,prl01}.
The interest on these topological objects,
however, has been confined mainly to theoretical physics, 
because most of them (exept the vortices) exist in
``hypothetical'' worlds which are very difficult to
create in laboratories. The only
topological objects which one can realistically expect to exist
in the ``standard'' models are the electroweak 
monopoles and dyons in Weinberg-Salam model
which have a non-trivial $W_\mu$
and $Z_\mu$ dressing \cite{plb97,yang}. Unfortunately these objects
could carry an infinite energy, which makes it impossible to
create them experimentally.

Fortunately the recent experimental advances in condensed matter
physics, in particular the construction of multi-component Bose-Einstein 
condensates (BEC) made of $^{87}{\rm Rb}$ \cite{bec,ruo}
and two-gap superconductors made of $MgB_2$ \cite{exp1,exp2}, 
have widely opened new opportunities for
us to create new topological objects experimentally which so far
have been only of theoretical interest. 
The purpose of this paper is to argue that these new 
multi-component condensates could allow us to have real
knots, topologically stable finite energy 3-dimensional solitons.
This is because, due to the multi-component structure, the new condensates
have a non-Abelian symmetry which provides the needed topology
for the stable knots.

To understand how the realistic knots can appear in these
condensed matters, it is necessary to understand 
the prototype Faddeev-Niemi knot in Skyrme theory. 
The Skyrme theory is well known to have a magnetic vortex 
known as the baby skyrmion \cite{piet}, and the
Faddeev-Niemi knot can be identified as a twisted vortex ring
made of a helical baby skyrmion \cite{prl01,plb04}. In the following
we show how one can construct similar knots from helical vortices 
in two-component Bose-Einstein condensates and two-gap superconductors. 

The paper is organized as follows. In Section II we discuss
how the helical vortex can give rise to the prototype knot 
in Skyrme theory for later purpose.
In Section III we discuss gauge theory of two-component BEC
which has the vorticity interaction, and show that the theory 
is closely related to Skyrme theory. With this we show that
the theory allows a topological knot similar to the one in 
Skyrme theory. We also identify that this knot is a vorticity knot 
very similar to the one in Gross-Pitaevskii theory of two-component BEC.
In Section IV we discuss the Landau-Ginzburg theory of two-gap
superconductor and show that the theory is closely related to 
the gauge theory of two-component BEC. With this we argue 
that the theory can also admit a knot, a twisted magnetic vortex 
ring, similar to the vorticity knot in two-component BEC.
In doing so we also establish the non-Abelian flux quantization
in two-gap superconductor. We demonstrate the existence of magnetic 
vortex whose flux is quantized in the unit $4\pi/g$, not $2\pi/g$, 
in two-gap superconductor.
Finally in Section V we discuss physical implications of 
our results.

\section{Knot in Skyrme theory}

The Skyrme theory has a rich topological structure.
It has skyrmion and baby skyrmion \cite{skyrme,piet}.
But recently it has been shown that the theory allows
a knotted soliton identical to
the Faddeev-Niemi knot in Skyrme-Faddeev
non-linear sigma model, which can be identified as a twisted 
vortex ring made of helical baby skyrmion \cite{prl01,plb04}. 
To see this, we review the knot in Skyrme theory first. 

Let $\omega$ and $\hn$ be the scalar field and 
the non-linear sigma field in Skyrme theory. With
\bea
&U = \exp (\omega \dfrac{\vec \sigma}{2i} \cdot \hat n)
= \cos \dfrac{\omega}{2} - i (\vec \sigma \cdot \hat n) 
\sin \dfrac{\omega}{2}, \nn\\
&L_\mu = U\partial_\mu U^{\dagger}, ~~~~~({\hat n}^2 = 1)
\eea
the Skyrme Lagrangian is expressed as
\bea
&{\cal L} = \dfrac{\mu^2}{4} {\rm tr} ~L_\mu^2 + \dfrac{\alpha}{32}
{\rm tr} \left( \left[ L_\mu, L_\nu \right] \right)^2  \nn \\
&= - \dfrac{\mu^2}{4} \Big[ \dfrac{1}{2} (\partial_\mu \omega)^2
+(1-\cos\omega)(\partial_\mu \hat n)^2 \Big]  \nn \\
& -\dfrac{\alpha}{16} \Big[ \dfrac{1-\cos\omega}{2} (\partial_\mu
\omega \partial_\nu \hat n - \partial_\nu \omega \partial_\mu \hat n)^2  \nn\\
& + (1-\cos\omega)^2 (\partial_\mu \hat n \times \partial_\nu \hat
n)^2 \Big].
\label{slag}
\eea
The equation of motion is given by
\bea
&\dfrac{\mu^2}{4} \left[ \partial^2 \omega -\sin\omega 
(\partial_\mu \hat n)^2 \right] \nn\\
&+\dfrac{\alpha}{32} \sin\omega (\partial_\mu
\omega \partial_\nu \hat n  -\partial_\nu \omega \partial_\mu \hat n)^2 \nn\\
&+\dfrac{\alpha}{8} (1-\cos\omega)
\partial_\mu \big[ (\partial_\mu
\omega \partial_\nu \hat n  -\partial_\nu \omega \partial_\mu \hat
n) \cdot \partial_\nu \hat n \big] \nn\\
& - \dfrac{\alpha}{8} (1-\cos\omega) \sin\omega (\partial_\mu \hat n \times
\partial_\nu \hat n)^2 =0, \nn \\
&\partial_\mu \Big\{ \dfrac{\mu^2}{4} (1-\cos\omega) \hat n \times
\partial_\mu \hat n \nn\\
& + \dfrac{\alpha}{16} (1-\cos\omega)
\big[ (\partial_\nu \omega)^2
\hat n \times \partial_\mu \hat n
-(\partial_\mu \omega \partial_\nu \omega) \hat n \times
\partial_\nu \hat n \big] \nn\\
&+\dfrac{\alpha}{8} (1-\cos\omega)^2
(\hat n \cdot \partial_\mu \hat n \times
\partial_\nu \hat n) \partial_\nu \hat n \Big\}=0.
\label{seq}
\eea
With the spherically symmetric ansatz
\bea
\omega = \omega (r),~~~~~\hat n = \hat r,
\eea
(3) is reduced to
\bea
&\dfrac{d^2 \omega}{dr^2} +\dfrac{2}{r} \dfrac{d\omega}{dr} -\dfrac{2
\sin\omega}{r^2}
-\dfrac{\alpha}{\mu^2} \Big[ \dfrac{\sin^2 (\omega /2)}{r^2}
\dfrac{d^2 \omega}{dr^2} \nn\\
&+\dfrac{\sin\omega}{2 r^2} (\dfrac{d\omega}{dr})^2
-\dfrac{2 \sin\omega \sin^2 (\omega /2)}{r^4} \Big] =0.
\eea
Imposing the boundary condition $\omega(0)= 2\pi$ and $\omega(\infty)= 0$,
one has the well-known skyrmion \cite{skyrme}.

A remarkable point of Skyrme theory is that $\omega=\pi$,
independent of $\hn$, becomes a classical solution \cite{prl01}.
So restricting $\omega$ to $\pi$, one can reduce the Lagrangian 
(\ref{slag}) to the Skyrme-Faddeev Lagrangian
\bea
{\cal L} \rightarrow -\dfrac{\mu^2}{2} (\partial_\mu
\hat n)^2-\dfrac{\alpha}{4}(\partial_\mu \hat n \times
\partial_\nu \hat n)^2,
\label{sflag}
\eea
in which case the equation of motion (\ref{seq}) is reduced to
\bea
&\hn \times
\partial^2 \hn + \dfrac{\alpha}{\mu^2} ( \partial_\mu N_{\mu\nu} )
\partial_\nu \hn = 0, \nn\\
&N_{\mu\nu} = \hn \cdot (\partial_\mu \hn \times \partial_\nu \hn)
=\pro_\mu C_\nu - \pro_\nu C_\mu.
\label{sfeq}
\eea
Notice that since $N_{\mu\nu}$ forms a closed two-form, 
it always admits a $U(1)$ potential $C_\mu$.

The equation (\ref{sfeq}) allows non-Abelian monopole,
baby skyrmion, helical baby skyrmion, and Faddeev-Niemi
knot as its solutions \cite{prl01,plb04}. But for our purpose it is 
important to understand the helical baby skyrmion, because this 
plays a crucial role for us to construct the knot. 
So we review the helical baby skyrmion.

\begin{figure}[t]
\includegraphics[scale=0.5]{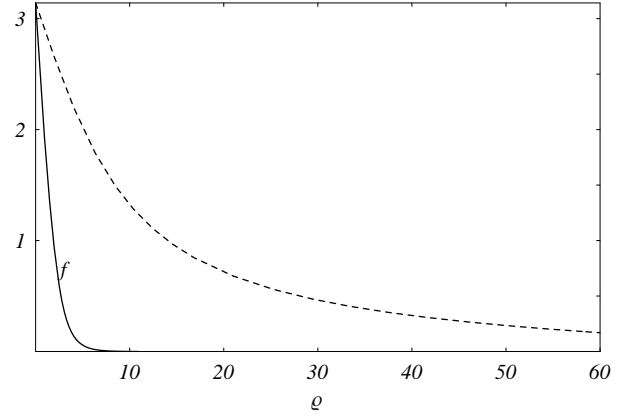}
\caption{The baby skyrmion (dashed line) with $m=0,n=1$
and the helical baby skyrmion (solid line) with $m=n=1$ in Skyrme
theory. Here $\varrho$ is in the unit ${\sqrt \alpha}/\mu$
and $k=0.8~\mu/{\sqrt \alpha}$.}
\label{hbs}
\end{figure}

To construct the desired helical baby skyrmion
let $(\varrho,\varphi,z)$ the cylindrical coodinates,
and choose the ansatz
\bea
&\n= \Bigg(\matrix{\sin{f(\varrho)}\cos{(mkz+n\varphi)} \cr
\sin{f(\varrho)}\sin{(mkz+n\varphi)} \cr \cos{f(\varrho)}}\Bigg).
\label{hvans}
\eea
With this the equation (\ref{sfeq}) is reduced to
\bea
&\Big(1+(m^2 k^2+\dfrac{n^2}{\varrho^2})
\dfrac{\sin^2{f}}{g^2 \rho^2}\Big) \ddot{f}
+ \Big( \dfrac{1}{\varrho}+ 2\dfrac{\dot{\rho}}{\rho} \nn\\
&+(m^2 k^2+\dfrac{n^2}{\varrho^2})
\dfrac{\sin{f}\cos{f}}{g^2 \rho^2} \dot{f}
+ \dfrac{1}{\varrho} (m^2 k^2-\dfrac{n^2}{\varrho^2})
\dfrac{\sin^2{f}}{g^2 \rho^2} \Big) \dot{f} \nn\\
&- (m^2 k^2+\dfrac{n^2}{\varrho^2}) \sin{f}\cos{f}=0.
\label{hveq}
\eea
So with the boundary condition
\bea
f(0)=\pi,~~f(\infty)=0,
\label{bc}
\eea
we obtain the non-Abelian vortex solutions shown in Fig.\ref{hbs}.
Notice that, when $m=0$, the solution describes
the well-known baby skyrmion \cite{piet}. But when $m$ is not zero,
it describes a helical baby skyrmion, a twisted magnetic vortex 
which is periodic in $z$-coodinate \cite{plb04}.
In this case, the vortex has a quantized magnetic flux
not only along the $z$-axis but also around the $z$-axis.

The helical baby skyrmion will become
unstable unless the periodicity condition is enforced by
hand. But it plays a very important role
because one can make it a vortex ring by smoothly connecting
two periodic ends. In this case the vortex ring 
acquires the topology of a knot, and thus becomes a knot 
itself \cite{prl01,plb04}. In fact it becomes a knot made of two
magnetic fluxes linked together, whose knot topology is described by
the Chern-Simon index of the potential $C_\mu$,
\bea
&Q = \dfrac{1}{32\pi^2} \int \epsilon_{ijk} C_i N_{jk} d^3x = mn,
\label{kqn}
\eea
which describes the Hopf mapping $\pi_3(S^2)$ defined by $\hn$.

Clearly the knot has a topological stability, because two
flux rings linked together can not be disjointed by a smooth
deformation of the field configuration.
Moreover the topological stability is 
backed up by a dynamical stability. This is because
the knot can be viewed as two magnetic fluxes linked together,
and the magnetic flux trapped in the knot disk
can not be squeezed out. This provides
a stablizing repulsive force which prevent the collapse 
of the knot \cite{plb04}.

From our discussion it becomes clear that the existence of 
the helical baby skyrmion is crucial for the existence of a topologically
stable knot. In the following we show that an identical mechanism 
works in two-component BEC and two-gap superconductor 
which guarantees the existence of a stable knot.

\section{Knot in Two-component Bose-Einstein Condensates}

The recent advent of multi-component BEC (in particular
the spin-1/2 condensate of $^{87}{\rm Rb}$ atoms)
has widely opened a new opportunity for
us to study novel topological objects which can not be
realized in ordinary (one-component) BEC \cite{bec,ruo}.
This is because the multi-component BEC naturally allows
a non-Abelian structure which accommodates a non-trivial
topological objects, in particular a topolgical knot which is
very similar to the knot in Skyrme theory \cite{bec1,bec5}.
 
There are two competing theories of two-component BEC,
the popular Gross-Pitaevskii theory \cite{ruo} and the gauge theory
of two-component BEC proposed recently \cite{bec1}. 
Both theories predict topological knots. 
Many authors have already claimed 
the existence of a knot in Gross-Pitaevskii theory \cite{ruo}.
Moreover it has been shown that this knot can be identified 
as a vorticity knot which is made of
two vorticity fluxes linked together, whose topology $\pi_3(S^2)$
is fixed by the Chern-Simon index of the velocity potential
of the condensate \cite{bec5}.

So in this section we discuss the gauge theory of two-component BEC,
and show that this theory also has a vorticity knot very similar 
to the knot in Gross-Pitaevskii theory. 
Consider a ``charged'' two-component BEC
described by a complex doublet $\phi$ interacting
``electromagnetically'', which can be described by the following
non-relativistic gauged Gross-Pitaevskii type Lagrangian (we
will discuss the relativistic generalization later)
\bea
&{\cal L} = i \dfrac {\hbar}{2} \Big[\phi^\dag (\tilde D_t \phi)
-(\tilde D_t \phi)^\dag \phi \Big] 
-  \dfrac {\hbar^2}{2m} |\tilde D_i \phi|^2 \nn\\
& + \mu^2 \phi^\dag \phi - \dfrac {\lambda}{2}
(\phi^\dag \phi)^2 - \dfrac {1}{4} \tilde F_{\mu \nu} ^2,
\label{beclag}
\eea
where $\tilde D_\mu \phi = ( \pro_\mu - i g \tilde A_\mu ) \phi$,
and $\mu^2$ and $\lambda$ are the chemical potential 
and the quartic coupling constant.
Normalizing $\phi$ to $(\sqrt{2m}/\hbar)\phi$ and putting
\bea
\phi = \dfrac {1}{\sqrt 2} \rho \xi , ~~~~~(\xi^\dag \xi = 1)
\eea
we have the following Hamiltonian in the static limit,
\bea
&{\cal H} =  \dfrac {1}{2} (\pro_i \rho)^2
+ \dfrac {\rho^2}{2} |\tilde D_i \xi |^2 \nn \\
&- \dfrac {\mu^2}{2} \rho^2 + \dfrac{\lambda}{8} \rho^4
+ \dfrac {1}{4} \tilde F_{ij} ^2.
\label{becham}
\eea
Notice that here we have also
normalized $\mu^2$ and $\lambda$ to
$(\hbar^2/2m) \mu^2$ and $(\hbar^2/2m)^2 \lambda$.
From now on we will use the normalized Hamiltonian (\ref{becham}).

At this point it is important to realize that the ``electromagnetic''
interaction here should be self-induced, since we are dealing
with neutral condensates. So we identify
the ``electromagnetic'' potential by the velocity field of
the doublet $\xi$,
\bea
\tilde A_\mu = -\dfrac {i}{g} \xi^\dag \pro_\mu \xi.
\label{am}
\eea
A justification for this is that the $U(1)$ gauge invariance almost
dictates us to identify the velocity field as the gauge potential.
Indeed in the absence of the Maxwell term, (\ref{am}) becomes 
an equation of motion. With this identification 
the field strength becomes the vorticity field
\bea
&\tilde F_{\mu\nu} = \pro_\mu \tilde A_\nu - \pro_\nu \tilde A_\mu \nn\\
&= -\dfrac {i}{g} (\pro_\mu \xi^\dag \pro_\nu \xi
- \pro_\nu \xi^\dag \pro_\mu \xi).
\label{fmn}
\eea
One might wonder why we have included 
the vorticity interaction in the Lagrangian (\ref{beclag}), 
when this is absent in the Gross-Pitaevskii Lagrangian. 
The reason is because it costs energy to create
the vorticity in two-component BEC. In ordinary BEC one does
not have to worry about the vorticity because the vorticity is identically
zero (since the velocity is given by the gradient of the phase
of the one-component complex condensate). But a non-Abelian
(multi-component) BEC has a non-vanishing vorticity,
in which case it is natural to include
the vorticity interaction in the Lagrangian \cite{bec1,bec5}. 

With (\ref{am}) the Hamiltonian (\ref{becham}) becomes
\bea
&{\cal H} = \dfrac {1}{2} (\pro_i \rho)^2 + \dfrac {\rho^2}{2}\Big(|\pro_i
\xi |^2 - |\xi^\dag \pro_i \xi|^2 \Big) 
- \dfrac {\lambda}{8} (\rho^2 - \rho^2_0)^2 \nn\\
&- \dfrac {1}{4 g^2} (\pro_i \xi^\dag \pro_j \xi 
- \pro_j \xi^\dag \pro_i \xi)^2, \nn\\
&\rho^2_0= \dfrac{2\mu^2}{\lambda}.
\label{becham1}
\eea
Notice that now the coupling constant $g$ represents
the coupling strength of the vorticity.
Minimizing the Hamiltonian we have
the following equation of motion
\bea
& \pro^2 \rho - \Big(|\pro_i \xi |^2 - |\xi^\dag \pro_i \xi|^2 \Big)
\rho = \dfrac {\lambda}{2} (\rho^2 - \rho_0^2) \rho,\nn \\
&\Big\{(\pro^2 - \xi^\dag \pro^2 \xi) + 2 (\dfrac {\pro_i
\rho}{\rho} - \xi^\dag \pro_i\xi)(\pro_i - \xi^\dag \pro_i \xi) \nn\\
&- \dfrac {1}{g^2 \rho^2} \Big[\pro_i (\pro_i
\xi^\dag \pro_j \xi - \pro_j \xi^\dag \pro_i \xi) \Big] 
(\pro_j - \xi^\dag \pro_j \xi) \Big\} \xi \nn\\
&= 0.
\label{beckeq1}
\eea
To understand the meaning of this notice that with
\bea
\n = \xi^\dag \vec \sigma \xi,
\label{hn}
\eea
we have
\bea
& |\pro_\mu \xi|^2 - |\xi^\dag \pro_\mu \xi|^2
= \dfrac{1}{4} (\pro_\mu \hn)^2, \nn\\
&i (\pro_\mu \xi^\dag \pro_\nu \xi 
- \pro_\nu \xi^\dag \pro_\mu \xi ) 
= \dfrac{1}{2} \hn \cdot (\pro_\mu \hn \times \pro_\nu \hn) \nn\\
&= \dfrac{1}{2} N_{\mu\nu},
\label{id}
\eea
where $N_{\mu\nu}$ is mathematically identical to what we have 
in Skyrme theory in (\ref{sfeq}).
This tells that the potential $C_\mu$ for the two-form $N_{\mu\nu}$
in (\ref{sfeq}) is given by (up to a gauge transformation)
\bea
C_\mu = 2g \tilde A_\mu = -2i \xi^\dag \pro_\mu \xi. 
\label{cm}
\eea
More importantly, with (\ref{id}) the Hamiltonian (\ref{becham1}) 
can be expressed as
\bea
&{\cal H} = \dfrac{1}{2} (\pro_i \rho)^2
+ \dfrac{\rho^2}{8} (\pro_i \hn)^2 
+ \dfrac{\lambda}{8}(\rho^2-\rho^2_0)^2 \nn\\
&+ \dfrac{1}{16 g^2} (\pro_i \hn \times \pro_j \hn)^2.
\eea
So the theory becomes very similar to Skyrme-Faddeev
theory, which strongly indicates that 
the gauge theory of two-component BEC can allow a knot.
As importantly this strongly implies that the Skyrme-Faddeev theory 
could play an important role in condensed matter
physics.

To simplify the equation (\ref{beckeq1}) notice that 
from (\ref{am}) and (\ref{hn}) we have
\bea
\pro_\mu \xi = (ig \tilde A_\mu + \dfrac{1}{2} \vec \sigma
\cdot \pro_\mu \n) \xi.
\eea
With the identity the second equation of (\ref{beckeq1}) is reduced to 
\bea
&(A + \B \cdot \vec \sigma) \xi = 0, \nn\\
&A = (\pro_i \n)^2, \nn \\
&\B = \pro^2 \n + 2 \dfrac{\pro_i \rho}{\rho} \pro_i \n
- \dfrac{i}{2 g^2 \rho^2} \pro_i N_{ij} \pro_j \n, 
\eea
which can be written as
\bea
&A+ \vec B \cdot \hn=0, \nn\\
& \hn \times \vec B - i \hn \times (\hn \times \vec B) =0,
\eea
or 
\bea
\n \times (\pro^2 \n + 2 \dfrac{\pro_i \rho}{\rho} \pro_i \n) 
+ \dfrac{1}{2 g^2 \rho^2} \pro_i N_{ij} \pro_j \n = 0.
\eea
So we can put (\ref{beckeq1}) into the form
\bea
&\pro^2 \rho - \dfrac{1}{4} (\pro_i \n)^2 \rho 
= \dfrac{\lambda}{2}(\rho^2 - \rho_0^2)
\rho, \nn \\
&\n \times \pro^2 \n + 2 \dfrac{\pro_i \rho}{\rho} \n \times \pro_i
\n + \dfrac{1}{g^2\rho^2} \pro_i N_{ij} \pro_j \n = 0.
\label{beckeq2}
\eea
This is the equation of two-component BEC that we are 
looking for. The similarity between this and the equation 
(\ref{sfeq}) of Skyrme-Faddeev theory is unmistakable.

Notice that the target space of $\xi$ and $\n$ is $S^3$ and
$S^2$, but here we have transformed the equation for $\xi$
in (\ref{beckeq1}) completely into the equation
for $\n$ in (\ref{beckeq2}).
This is made possible because, with the Abelian gauge invariance of (6),
the physical target space of $\xi$ becomes the gauge orbit space
$S^2 = S^3/S^1$ which forms a $CP^1$ space.
This means that we can view the theory as a self interacting
$CP^1$ theory (coupled to a scalar field $\rho$), and replace
$\xi$ completely in terms of $\n$.

\begin{figure}[t]
\includegraphics[scale=0.5]{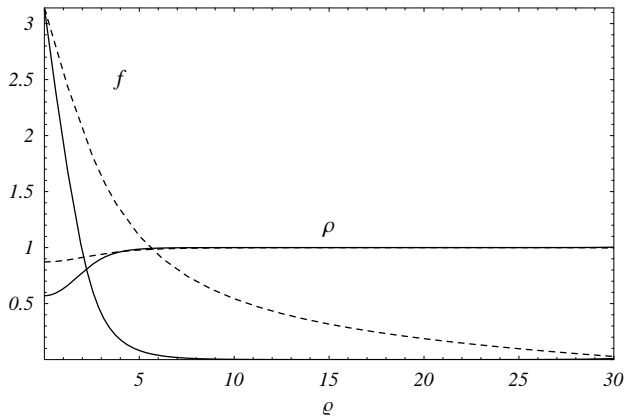}
\caption{The non-Abelian vortex (dashed line) with $m=0,n=1$ and
the helical vortex (solid line) with $m=n=1$ in the gauge theory
of two-component BEC. Here we have put $\lambda/g^2=1$,
$k=0.64~{\sqrt \lambda}\rho_0$, and $\varrho$ is in the unit of
$1/{\sqrt \lambda}\rho_0$.}
\label{becheli}
\end{figure}

To show that the theory has a knot solution we construct 
a helical vortex solution first. To do this 
we choose the ansatz
\bea
&\rho= \rho(\varrho), 
~~~~~\xi = \Bigg( \matrix{\cos \dfrac{f(\varrho)}{2}
\exp (-in\varphi) \cr
\sin \dfrac{f(\varrho)}{2} \exp (imkz)} \Bigg), \nn\\
&\n= \xi^\dag \vec \sigma \xi
= \Bigg(\matrix{\sin{f(\varrho)}\cos{(mkz+n\varphi)} \cr
\sin{f(\varrho)}\sin{(mkz+n\varphi)} \cr \cos{f(\varrho)}}\Bigg). \nn\\
\label{bhvans}
\eea
With this (\ref{beckeq2}) is reduced to
\bea
&\ddot{\rho}+\dfrac{1}{\varrho}\dot\rho
- \dfrac{1}{4}\Big(\dot{f}^2+(m^2 k^2+\dfrac{n^2}{\varrho^2})
\sin^2{f}\Big)\rho \nn\\
&= \dfrac{\lambda}{2}(\rho^2-\rho_0^2)\rho, \nn\\
&\Big(1+(m^2 k^2+\dfrac{n^2}{\varrho^2})
\dfrac{\sin^2{f}}{g^2 \rho^2}\Big) \ddot{f}
+ \Big( \dfrac{1}{\varrho}+ 2\dfrac{\dot{\rho}}{\rho} \nn\\
&+(m^2 k^2+\dfrac{n^2}{\varrho^2})
\dfrac{\sin{f}\cos{f}}{g^2 \rho^2} \dot{f}
+ \dfrac{1}{\varrho} (m^2 k^2-\dfrac{n^2}{\varrho^2})
\dfrac{\sin^2{f}}{g^2 \rho^2} \Big) \dot{f} \nn\\
&- (m^2 k^2+\dfrac{n^2}{\varrho^2}) \sin{f}\cos{f}=0.
\label{bveq}
\eea
So with the boundary condition
\bea
&\dot \rho(0)=0,~~~~~\rho(\infty)=\rho_0, \nn\\
&f(0)=\pi,~~~~~f(\infty)=0,
\label{becbc}
\eea
we obtain the non-Abelian vortex solutions shown in Fig.\ref{becheli}.

There are three points that have to be
emphasized here. First, when $m=0$, the solution describes
the untwisted non-Abelian vortex \cite{bec1}. But when $m$ is not zero,
it describes a helical vortex which is periodic in $z$-coodinate.
In this case, the vortex has a non-vanishing velocity current
(not only around the vortex but also) along the $z$-axis.
Secondly, the doublet $\phi$ starts from
the second component at the core,
but the first component takes over completely at the infinity.
This is due to the boundary condition $f(0)=\pi$ and $f(\infty)=0$,
which assures that our solution describes a genuine
non-Abelian vortex. Thirdly, when $f=0$ (or $f=\pi$)
the doublet effectively becomes
a singlet, and (\ref{bveq}) describes the well-known
vortex in single-component BEC. Only when $f$ has a non-trivial
profile, we have a non-Abelian vortex.

In Skyrme theory the helical vortex is interpreted as
a twisted magnetic vortex whose flux is quantized \cite{prl01,plb04}
Now we show that the above vortex is a twisted vorticity flux
which is also quantized.
To see this notice that the non-Abelian structure of
the vortex is represented by the doublet $\xi$.
Moreover, the velocity field 
of the doublet is given by 
\bea
&\tilde A_\mu =-\dfrac{n}{2g}(\cos{f}+1) \pro_\mu \varphi \nn\\
&-\dfrac{mk}{2g}(\cos{f}-1) \pro_\mu z,
\label{gpvel}
\eea
which generates the vorticity
\bea
&H_{\mu\nu}= -\dfrac{i}{g}(\pro_\mu \xi^{\dagger} \pro_\nu \xi
-\pro_\nu \xi^{\dagger} \pro_\mu \xi) \nn\\
&=\dfrac{\dot{f}}{2g} \sin{f}\Big(n(\pro_\mu \varrho \pro_\nu \varphi
-\pro_\nu \varrho \pro_\mu \varphi) \nn\\
&+mk(\pro_\mu \varrho \pro_\nu z
- \pro_\nu \varrho \pro_\mu z) \Big).
\label{gpvor}
\eea
This has two vorticity fluxes, $\phi_{\hat z}$
along the $z$-axis
\bea
&\phi_{\hat z}=\dfrac{}{} \int H_{{\hat \varrho}{\hat \varphi}}
\varrho d \varrho d \varphi
= -\dfrac{2\pi n}{g},
\label{gpfluxz}
\eea
and $\phi_{\hat \varphi}$ around the the $z$-axis (in one
period section from $z=0$ to $z=2\pi/k$)
\bea
&\phi_{\hat \varphi}=\dfrac{}{} \int_{z=0}^{z=2\pi/k}
H_{{\hat z}{\hat \varrho}} d \varrho dz
= \dfrac{2\pi m}{g}.
\label{gpfluxphi}
\eea
This shows that the helical vortex is made of two quantized vorticity
fluxes, the $\phi_{\hat z}$-flux which is concentrated at the core
and the $\phi_{\hat \varphi}$-flux which surrounds it. 
This confirms that the helical vortex is a twisted vorticity
flux which is very similar to the helical vortex in Gross-Pitaevskii 
theory \cite{bec5}.

Now, with the helical vortex, one can easily make a twisted vortex ring
smoothly connecting the periodic ends together. 
And just as in Skyrme theory the twisted vortex ring
becomes a topological knot. But
here it is $\pi_3(S^3)$ of the doublet $\xi$ which provides the
non-trivial quantum number, 
\bea
q = - \dfrac {1}{4\pi^2} \int \epsilon_{ijk} \xi^{\dagger}
\partial_i \xi ( \partial_j \xi^{\dagger}
\partial_k \xi ) d^3 x.
\label{beckqn}
\eea
Of course this is identical to the expression (\ref{kqn}), due to the
Hopf fibering of $S^3$ to $S^2 \times S^1$ \cite{bec5}. This
tells that we can express the knot quantum number either
by $\pi_3(S^3)$ or by $\pi_3(S^2)$. 

Obviously this knot has a topological stability, because 
the knot topology (\ref{kqn}) can not be changed by a smooth 
deformation of the field configuration. Moreover it has 
a dynamical stability. To understand this notice that
the knot has a twisted velocity field so that it has a
non-vanishing velocity around the $z$-axis.
This means that it carries a non-vanishing angular momentum along 
the $z$-axis. And this angular momentum provides 
the dynamical stability, because it creates a centrifugal force that 
prevents the collapse of the knot. Notice that this dynamical 
stability originates from the knot topology, because 
the angular momentum comes from the twisted velocity field. 
In this sense the topological stability and the dynamical stability
have one and the same origin. It is this remarkable 
interplay between topology and dynamics which assures 
the stability of the knot. The nontrivial twisted topology of 
the knot expresses itself in the form of the angular momentum, 
which in turn provides the dynamical stability of the knot. 
This presence of the angular momnetum is 
what differentiates our
knot from the untwisted Abrikosov-type vortex ring 
which has no dynamical stability.  

There have been assertions that two-component BEC admits
a knot \cite{ruo}. But notice that this knot is based on 
Gross-Pitaevskii theory of two-component BEC, which has no
vorticity interaction. In contrast our knot is based on 
the gauge theory in which the vorticity 
interaction plays a crucial role. Nevertheless physically two knots 
are very similar \cite{bec5}. Both can be identified as 
a vorticity knot. This implies that both theories should be 
taken seriously as a theory of two-component BEC.

\section{Knot in Two-gap Superconductors}

In the above gauge theory of spin-1/2   
condensates the gauge interaction was a  
self-induced interaction. But when the doublet is charged,
the gauge interaction can be treated as independent.
In this case the theory can describe a two-gap superconductor.
But even in this case the knot topology and thus the knot itself
should survive. This implies that two-gap 
superconductor should also have
a topological knot.

The knot in two-gap superconductor could be either relativistic
or non-relativistic, and appear in both Abelian and non-Abelian
setting \cite{cm2}. In this paper we will discuss the relativistic 
knot in the Abelian setting (a non-relativistic
Gross-Pitaevskii type theory gives an identical result).
Consider a charged doublet scalar field $\phi$ coupled to
the real electromagnetic field,
\bea
&{\cal L} = - |D_\mu \phi|^2 + \mu^2 \phi^{\dagger}\phi 
- \dfrac{\lambda}{2} (\phi^{\dagger} \phi)^2
- \dfrac{1}{4} F_{\mu \nu}^2 , \nn\\
&D_\mu \phi = (\partial_\mu - ig A_\mu) \phi.
\label{sclag}
\eea
The Lagrangian has the equation of motion
\bea
&D^2\phi =\lambda(\phi^{\dagger} \phi
-\dfrac{\mu^2}{\lambda})\phi, \nn\\
&\partial_\mu F_{\mu \nu} = j_\nu =  i g \Big[(D_\nu
\phi)^{\dagger}\phi - \phi ^{\dagger}(D_\nu \phi) \Big].
\label{sceq1}
\eea
Now, with
\bea
\phi =\dfrac{1}{\sqrt 2} \rho \xi,~~~~~{\xi}^{\dagger}\xi = 1,
~~~~~\hat n = \xi^{\dagger} \vec \sigma \xi,
\eea
we can reduce (\ref{sceq1}) to \cite{cm2}
\bea
&\partial ^2 \rho - \Big(
\dfrac{1}{4} (\partial _\mu \hat n)^2+ g^2
(A_\mu + \tilde A_\mu)^2 \Big) \rho
= \dfrac{\lambda}{2} (\rho^2-\rho_0^2)\rho, \nn\\
&\hat n \times \partial ^2 \hat n + 2 \dfrac{\partial_\mu \rho}{
\rho} \hat n \times \partial_\mu \hat n
+ \dfrac{2}{g\rho^2} \partial_\mu F_{\mu\nu} \partial_\nu \hat n =0, \nn\\
&\partial_\mu F_{\mu\nu} = j_\mu
=g^2 \rho^2 (A_\mu + \tilde A_\mu), \nn\\
&\tilde A_\mu=-\dfrac{i}{g} \xi^{\dagger}\partial_\mu \xi,
~~~~~\rho_0=\dfrac{2\mu^2}{\lambda}.
\label{sceq2}
\eea
This is the equation for two-gap superconductor.
Notice that with $A_\mu=-\tilde A_\mu$ the first two equations
reduce to (\ref{beckeq2}). This tells that the gauge theory of 
two-component BEC and the above theory of two-gap superconductor
are closely related.

To obtain the desired knot we first construct a superconducting
helical magnetic vortex. Let 
\bea
&\rho=\rho(\varrho), 
~~~~~\xi = \Bigg( \matrix{\cos \dfrac{f(\varrho)}{2} \exp (-in\varphi) \cr
\sin \dfrac{f(\varrho)}{2} \exp (imkz)} \Bigg), \nn\\
&A_\mu= \dfrac{1}{g} \big(n A_1(\varrho) \partial_\mu\varphi
+ mk A_2(\varrho) \partial_\mu z \big), \nn\\
&\n= \xi^\dag \vec \sigma \xi
= \Bigg(\matrix{\sin{f(\varrho)}\cos{(n\varphi+mkz)} \cr
\sin{f(\varrho)}\sin{(n\varphi+mkz)} \cr \cos{f(\varrho)}}\Bigg), \nn\\
&\tilde A_\mu=-\dfrac{n}{2g} \big(\cos{f(\varrho)}+1\big)
\partial_\mu \varphi \nn\\
&- \dfrac{mk}{2g} \big(\cos{f(\varrho)}-1\big) \partial_\mu z.
\label{scans}
\eea
With this we have
\bea
&j_\mu = g\rho^2 \Big(n \big(A_1-\dfrac{\cos{f}+1}{2}\big)
\partial_\mu \varphi \nn\\
&+ mk\big(A_2-\dfrac{\cos{f}-1}{2}\big)
\partial_\mu z \Big),
\label{sc}
\eea
and (\ref{sceq2}) becomes
\bea
&\ddot{\rho}+\dfrac{1}{\varrho}\dot\rho
- \Big[\dfrac{1}{4} \Big(\dot{f}^2
+\big(\dfrac{n^2}{\varrho^2} + m^2 k^2 \big) \sin^2{f}\Big) \nn\\
&+\dfrac{n^2}{\varrho^2} \Big(A_1-\dfrac{\cos{f}+1}{2}\Big)^2
+ m^2 k^2 \Big(A_2-\dfrac{\cos{f}-1}{2}\Big)^2 \Big]\rho \nn\\
&= \dfrac{\lambda}{2}(\rho^2-\rho_0^2)\rho, \nn\\
&\ddot{f} + \big(\dfrac{1}{\varrho}
+2\dfrac{\dot{\rho}}{\rho} \big)\dot{f}
-2 \Big(\dfrac{n^2}{\varrho^2}\big(A_1-\dfrac{1}{2}\big) \nn\\
&+ m^2 k^2 \big(A_2+\dfrac{1}{2}\big) \Big)\sin{f} = 0, \nn\\
&\ddot{A}_1-\dfrac{1}{\varrho}\dot{A}_1 -g^2 \rho^2
\Big(A_1-\dfrac{\cos{f}+1}{2}\Big) = 0, \nn\\
&\ddot{A}_2+\dfrac{1}{\varrho}\dot{A}_2 -g^2 \rho^2
\Big(A_2-\dfrac{\cos{f}-1}{2}\Big) = 0.
\label{sceq3}
\eea
Now, we impose the following boundary condition for the non-Abelian
vortices \cite{cm2},
\bea
&\rho (0) = 0,~~~\rho(\infty) = \rho_0, 
~~~f (0) = \pi,~~~f (\infty) = 0, \nn\\
& A_1 (0) = -1,~~~A_1 (\infty) =1.
\label{scbc}
\eea
This need some explanation, because the boundary
value $A_1(0)$ is chosen to be $-1$, not $0$. This is to
assure the smoothness of $\rho(\varrho)$ and $f(\varrho)$
at the origin. Only with this boundary value they
become analytic at the origin.
At this point one might object the boundary condition, because it creates
an apparent singularity in the gauge potential at the origin.
But notice that this singularity is an unphysical
(coordinate) singularity which can easily be removed by a gauge transformation.
In fact the singularity disappears with the gauge transformation
\bea
\phi \rightarrow \phi \exp(in\varphi),
~~~~A_\mu \rightarrow A_\mu + \dfrac{n}{g} \partial_\mu \varphi,
\eea
which changes the boundary condition $A_1(0)=-1$ and 
$A_1(\infty)=1$ to $A_1(0)=0$ and $A_1(\infty) =2$.
Mathematically this boundary condition
has a deep origin, which has to do with the fact
that the Abelian $U(1)$ runs from $0$ to $2\pi$, but the $S^1$
fiber of $SU(2)$ runs from $0$ to $4\pi$ \cite{cm2}.
As for $A_2(\varrho)$, we choose $A_2(\infty)=0$ 
to make the supercurrent vanishing at infinity
and require the vortex superconducting. As we will see,
this requires a logarithmic divergence for $A_2(0)$.
The boundary condition will
have an important consequence in the following.

\begin{figure}
\includegraphics[scale=0.7]{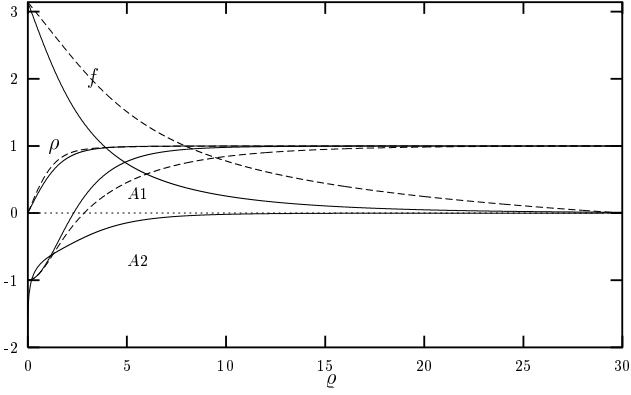}
\caption{The non-Abelian vortex (dashed line) with
$m=0,n=1$ and the helical vortex (solid line) with $m=n=1$ in
two-gap superconductor. Here we have put $g=1,~\lambda=2$,
$k=\rho_0/10$, and $\varrho$ is in the unit of $1/\rho_0$.
Notice that $A_2$ has a logarithmic singularity at the origin.}
\label{heli-sc}
\end{figure}

With the boundary condition we can integrate (\ref{sceq3})
and obtain the non-Abelian vortex solution
of the two-gap superconductor, which is shown in Fig.\ref{heli-sc}.
The solution is very similar to the one we have in 
two-component BEC. When $m=0$, the solution (with $A_2=0$) 
describes an untwisted non-Abelian vortex \cite{cm2}. 
But when $m$ is not zero, it describes a helical magnetic vortex 
which is periodic in $z$-coordinate.
Moreover, the vortex starts from
the second component at the core,
but the first component takes over completely at the infinity.
This is due to the boundary condition $f(0)=\pi$ and $f(\infty)=0$,
which assures that our solution describes a genuine
non-Abelian vortex. This is true even when $m=0$. Only when 
$f=0$ (or $f=\pi$) the doublet effectively becomes
a singlet, and (\ref{sceq3}) describes the Abelian
Abrikosov vortex of one-gap superconductor.

There are important differences between the non-Abelian vortex 
and the Abrikosov vortex. First the non-Abelian vortex
has a non-Abelian magnetic flux quantization \cite{cm2}. 
Indeed the quantized magnetic flux $\hat \phi_z$ 
of the non-Abelian vortex along the $z$-axis is given by 
\bea
&H_z= \dfrac{n}{g} \dfrac{\dot A_1}{\varrho}, \nn\\
&\hat \phi_z = \dfrac{}{} \int H_z d^2x
= \dfrac{2\pi n}{g} \big[A_1(\infty) - A_1(0)\big] \nn\\
&= \dfrac{4\pi n}{g}.
\label{zflux}
\eea
Notice that the unit of the non-Abelian flux is $4\pi/g$,
not $2\pi/g$. This is a direct
consequence of the boundary condition (\ref{scbc}).
This non-Abelian quantization of magnetic flux comes
from the non-Abelian topology $\pi_2(S^2)$ of the doublet $\xi$,
or equivalently the triplet $\hn$, whose topological
quantum number is given by \cite{cm2}
\bea
&q = \dfrac {g}{4\pi } \int H_z d^2 x
= - \dfrac {1}{4\pi} \int \epsilon_{ij} \partial_i \xi^{\dagger}
\partial_j \xi d^2 x \nn\\
&= \dfrac {1}{8\pi} \int \epsilon_{ij} \hn \cdot (\partial_i \hn
\times \partial_j \hn) d^2 x = n.
\label{scqn}
\eea
This distinguishes our non-Abelian vortex from
the Abelian vortex whose topology is fixed by $\pi_1(S^1)$.

Another important feature of the non-Abelian vortex 
is that it carries a non-vanishing supercurrent
along the $z$-axis,
\bea
&i_z = mkg \dfrac{}{} \int \rho^2 \big(A_2
-\dfrac{\cos f +1}{2}\big)\varrho d\varrho d\varphi \nn\\
&= \dfrac{2\pi mk}{g} \int \big(\ddot A_2
+\dfrac{1}{\varrho} \dot A_2 \big) \varrho d\varrho \nn\\
&= \dfrac{2\pi mk}{g} (\varrho \dot A_2) \Big|_{\varrho=0}^{\varrho=\infty}
=-\dfrac{2\pi mk}{g} (\varrho \dot A_2) \Big|_{\varrho=0}.
\label{scz}
\eea
This is due to the logarithmic divergence of $A_2$ at the origin.

Notice that the superconducting helical vortex
has only a heuristic value, because one needs
an infinite energy to create it (since the magnetic flux
around the vortex becomes divergent because of the singularity
of $A_2$ at the origin).
With the helical vortex, however, one can make
a vortex ring by smoothly bending and connecting
two periodic ends. In the vortex ring the infinite magnetic
flux of $A_2$ can be made finite making the finite supercurrent
(\ref{scz}) of the vortex ring produce a finite flux,
and we can fix the flux to have the value $4\pi m/g$
by adjusting the current with
$k$. With this the vortex ring now
becomes a topologically stable knot.

To see this notice that the doublet $\xi$, after forming a knot,
acquires a non-trivial topology $\pi_3(S^2)$ which provides
the knot quantum number,
\bea
&Q = - \dfrac {1}{4\pi^2} \int \epsilon_{ijk} \xi^{\dagger}
\partial_i \xi ( \partial_j \xi^{\dagger}
\partial_k \xi ) d^3 x \nn\\
&= \dfrac{g^2}{32\pi^2} \int \epsilon_{ijk} C_i
(\partial_j C_k - \partial_k C_j) d^3x
=mn.
\label{bkqn}
\eea
This is nothing but the Chern-Simon index of the potential
$C_\mu$, which is mathematically identical to
the quantum number of the knots we discussed before. 
This tells that our
knot is also made of two quantized magnetic flux rings
linked together whose knot quantum number is fixed by
the linking number $mn$. Obviously two flux rings linked together
can not be separated by any continuous deformation of
the field configuration. This provides the topological stability
of the knot.

Again this topological stability is backed up
by a dynamical stability. To see this notice that the supercurrent
of the knot has two components, the one around
the knot tube which confines the magnetic flux along the knot,
but more importantly the other along the knot
which creates a magnetic flux
passing through the knot disk. This component of supercurrent
along the knot now generates a net
angular momentum which provides
the centrifugal repulsive force preventing the knot to collapse.
This makes the knot dynamically stable.

To compare our knot with the Abrikosov vortex ring (made of the
Abrikosov vortex in conventional superconductor),
notice that the Abrikosov knot is empty (i.e.,
does not carry a net supercurrent).
As importantly it is unstable, and collapses immediately.
In contrast our knot has a helical supercurrent, and is stable. 
Furthermore these two features are deeply related. 
The helical supercurrent plays a crucial role to stablize the vortex
ring by providing the net angular momentum, which prevents the
collapse of the vortex ring. And this helical supercurrent 
originates from the knot topology. This remarkable
interplay between topology and dynamics is what provides
the stability of the knot. The nontrivial topology expresses 
itself in the form of the helical supercurrent, which in turn
provides the dynamical stability of the knot. We emphasize that
this supercurrent is what distinguishes our knot from
the Abrikosov vortex ring, which has neither topological
nor dynamical stability.

\section{Discussion}

\begin{figure}[t]
\includegraphics[scale=0.7]{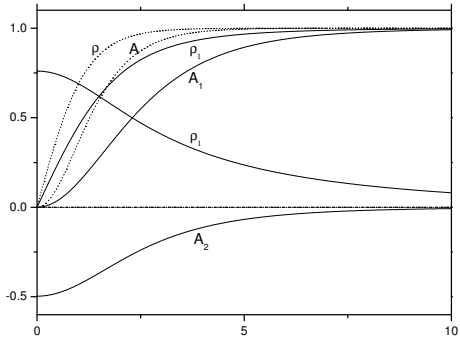}
\caption{The regular helical magnetic vortex with $m=n=1$ in
two-gap superconductor. Here we have put $g=1,~\lambda=2$,
$k=0.12 \rho_0$, $A_2(0)=0.5$, and $\varrho$ is in the unit of $1/\rho_0$.
Notice that the solution is completely regular.}
\label{2schv}
\end{figure}

In this paper we have presented a compelling argument
for the existence of topological
knots in two-component BEC and two-gap superconductor. 
Similar knots have popped out almost everywhere,
in particular in high energy physics in QCD \cite{plb05} and 
Weinberg and Salam model \cite{ewknot}. 
But we emphasize that at the center of these
topological objects lies the baby skyrmion and the Faddeev-Niemi
knot \cite{prl01,plb04}. In fact, our helical vortices 
and knots in this paper are a straightforward
generalization of the baby skyrmion and the Faddeev-Niemi
knot. 

It has been assumed that the topological
objects in Skyrme theory can only be realized at high energy,
at the hadronic scale. But our analysis shows that
similar objects could exist in a completely different environment,
at a much lower scale, in low energy condensed matters.
If so, the challenge now is
to verify the existence of the topological
knot experimentally in condensed matters. Constructing the knot
might not be so easy at present moment. 
Nevertheless, with some experimental ingenuity,
one should be able to construct the knots
in condensed matters.

Note Added: One might doubt the existence of a superconducting
knot because the superconducting helical vortex we discussed 
in Section IV was singular (and thus unphysical). In this note 
we report a regular superconducting helical vortex which has a finite 
magnetic flux around the axis and thus a finite energy. 
The regular solution is obtained linking $A_2(0)$ with $k$. 
For example for $k=0.12$ we obtain the regular solution shown 
in Fig.\ref{2schv}, with $A_2(0)=0.5$. This type of regular
helical vortex has vanishing supercurrent $i_z$, 
but could still be called superconducting 
because it has a non-trivial supercurrent density $j_z$
which generates a net magnetic flux $H_\varphi$ around the vortex.
The regular helical vortex strongly support the existence of
a regular knot. The details will be published elsewhere.
 
{\bf ACKNOWLEDGEMENT}

~~~The work is supported in part by the ABRL Program of
Korea Science and Engineering Foundation (R14-2003-012-01002-0)
and by the BK21 Project of the Ministry of Education.

\end{document}